\documentclass[prd,onecolumn]{revtex4}
\usepackage{dcolumn}
\usepackage{multirow}
\usepackage{graphicx}
\usepackage{amssymb}
\usepackage{bm}
\usepackage{hyperref}%for .pdf
\usepackage{epstopdf}%for .pdf
\usepackage{color}
\usepackage{mathrsfs}
\usepackage{amsmath,amssymb,amsthm}
\usepackage{rotating}
\usepackage{sverb, longtable}
\usepackage{subfigure}

\usepackage[graphicx]{realboxes}
\usepackage{adjustbox}
\begin{document}

\title{Pantheon+ tomography and Hubble tension}
%Pantheon+ tomography and $H_0$ tension
\author{Deng Wang}
\email{cstar@nao.cas.cn}
\affiliation{Instituto de F\'{i}sica Corpuscular (CSIC-Universitat de Val\`{e}ncia), E-46980 Paterna, Spain \\
	National Astronomical Observatories, Chinese Academy of Sciences, Beijing, 100012, China}
\begin{abstract}
The recently released Type Ia supernovae (SNe Ia) sample, Pantheon+, is an updated version of Pantheon and has very important cosmological implications. To explore the origin of the enhanced constraining power and internal correlations of datasets in different redshifts, we perform a comprehensively tomographic analysis of the Pantheon+ sample without and with the Cepheid host distance calibration, respectively. Specifically, we take two binning methods to analyze the Pantheon+ sample, i.e., equal redshift interval and equal supernovae number for each bin. For the case of equal redshift interval, after dividing the sample to 10 bins, the first bin in the redshift range $z\in[0.00122, \, 0.227235]$ dominates the constraining power of the whole sample. For the case of equal supernovae number, the first three low redshift bins prefer a large matter fraction $\Omega_m$ and only the sixth bin gives a relatively low cosmic expansion rate $H_0$. For both binning methods, we find no obvious evidence of evolution of $H_0$ and $\Omega_m$ at the $2\,\sigma$ confidence level. The inclusion of the SH0ES calibration can significantly compress the parameter space of background dynamics of the universe in each bin. When not considering the calibration, combining the Pantheon+ sample with cosmic microwave background, baryon acoustic oscillations, cosmic chronometers, galaxy clustering and weak lensing data, we give the strongest $1\,\sigma$ constraint $H_0=67.88\pm0.42$ km s$^{-1}$ Mpc$^{-1}$. However, the addition of the calibration leads to a global shift of the parameter space from the combined constraint and $H_0=68.66\pm0.42$ km s$^{-1}$ Mpc$^{-1}$, which is inconsistent with the Planck-2018 result at about $2\,\sigma$ confidence level.

\end{abstract}
\maketitle

\section{Introduction}
So far, in modern cosmology, there are two main problems about the background evolution of the universe, i.e., the nature of dark energy (DE) and the Hubble constant ($H_0$) tension. Further understanding and explorations of them will be very important for possible new physics. It has been almost a quarter of decade since DE is discovered independently by two SNe Ia search teams \cite{SupernovaCosmologyProject:1998vns,SupernovaSearchTeam:1998fmf}. However, the nature of DE is still unclear and intriguing. Currently, we just know its serveral basic properties \cite{DiValentino:2020vhf}: (i) DE is homogeneously permeated in the universe on cosmic scales and can hardly cluster unlike the dark matter (DM); (ii) DE behaves as a phenomenological fluid with equation of state (EoS) $\omega\approx-1$. Interestingly, the nature of DE is closely related to the severe $H_0$ tension, which states that the globally derived $H_0$ value from cosmic microwave background (CMB) under the assumption of $\Lambda$CDM \cite{Planck:2018vyg} is $5\,\sigma$ lower than the locally direct measurement of today's cosmic expansion rate from the Hubble Space Telescope (HST) \cite{Riess:2021jrx}. Maybe the answer to the question what DE is actually at all can be acquired during the process of solving the $H_0$ tension. Currently, there are a large number of models to relieve or even solve the $H_0$ discrepancy (see Refs.\cite{Abdalla:2022yfr,DiValentino:2020zio,Verde:2019ivm,Knox:2019rjx,Jedamzik:2020zmd,DiValentino:2021izs,Perivolaropoulos:2021jda,Shah:2021onj,Kamionkowski:2022pkx} for recent reviews).

Besides constructing a physical model to address the above two problems, another important approach is studying them by using different cosmological and astrophysical observations. To alleviate the $H_0$ tension, one may attempt to identify the unknown systematic uncertainties in local HST observations or determine $H_0$ with different datasets \cite{Abdalla:2022yfr,DiValentino:2020zio,Verde:2019ivm,Knox:2019rjx,Jedamzik:2020zmd,DiValentino:2021izs,Perivolaropoulos:2021jda,Shah:2021onj,Kamionkowski:2022pkx}. Traditionally, in order to beak the parameter degeneracy, one often implements the constraints on cosmological parameters by combining different probes together. For instance, the Planck collaboration obtains a tight constraint on DE EoS by combining CMB with baryon acoustic oscillations (BAO) and SNe Ia observations \cite{Planck:2018vyg}. Nonetheless, an elegant method is using an independent and powerful probe to study the evolutional behaviors of DE and $H_0$ problem. Although combined observations can reduce statistical errors to a large extent, unknown uncertainties and complexities may emerge. As the discovery tool of DE, SNe Ia has a strong potential to help probe the background dynamics of the universe. 

About eight years ago, the first integrated SNe Ia sample is the ``Joint Light-curve Analysis'' (JLA) constructed from the Supernova Legacy Survey (SNLS) and Sloan Digital Sky Survey (SDSS), which consists of 740 SNe Ia covering the redshift range $z\in(0.01, \, 1.3)$ \cite{SDSS:2014iwm}. In 2018, the second compilation is the Pantheon sample consisting of 1048 SNe Ia covering the redshift range $z\in(0.01, \, 2.3)$, which is made of 276 Pan-STARRS1 SNe Ia with useful distance estimations of SNe Ia from SNLS, SDSS, low-z and HST observations \cite{Pan-STARRS1:2017jku}. Recently, Refs.\cite{Scolnic:2021amr,Brout:2022vxf} release a new sample called Pantheon+, which is made of 1701 light curves of 1550 spectroscopically confirmed SNe Ia coming from 18 different sky surveys. This larger sample has a significant increase at low redshifts and lies in the redshift range $z\in[0.00122, \, 2.26137]$. We notice that the authors in Ref.\cite{Brout:2022vxf} give a joint constraint on $\Lambda$CDM from the Pantheon+ and HST $H_0$ measurement, but they do not provide a careful cosmological analysis of Pantheon+ at different redshift bins. As a consequence, in this study, we take a tomographic analysis of Pantheon+ sample to constrain $\Lambda$CDM, and study the effects of different redshift bins on constrained $H_0$ and present-day matter fraction $\Omega_{m}$. We take two binning methods to analyze the Pantheon+ sample, i.e., equal redshift interval and equal supernovae number for each bin.
For the former method, we find that the first bin in the redshift range $z\in[0.00122, \, 0.227235]$ dominates the constraining power of the full sample when dividing the full sample to 10 bins. The inclusion of the Cepheid host distance calibration can obviously compress the parameter space of background dynamics of the universe and reduces the constrained values of $\Omega_m$ in each bin. For the latter method, the first three low redshift bins prefer a large matter fraction $\Omega_m$ and only the sixth bin gives a relatively low cosmic expansion rate. For both binning methods, we find no obvious evidence of evolution of $H_0$ and $\Omega_m$ at the $2\,\sigma$ confidence level.    

This work is structured as follows. In Section II, we introduce the basic formula of $\Lambda$CDM. In Section III, we describe the Pantheon+ data and our analysis methodology. In Section IV, we display the numerical results. Discussions and conclusions are presented in the final section.

\section{Basic formula}
The action of general relativity (GR) reads as 
\begin{equation}
S=\int d^4x\sqrt{-g}\left[R-2\Lambda+\mathcal{L}_m\right], \label{1}
\end{equation}
where $g$, $R$, $\Lambda$ and $\mathcal{L}_m$ denote the trace of the spacetime metric, Ricci scalar, cosmological constant, standard matter Lagrangian, respectively. Varying Eq.(\ref{1}), we obtain the well-known Einstein field equation as 
\begin{equation}
R_{\mu\nu}-\frac{1}{2}g_{\mu\nu}R+\Lambda g_{\mu\nu}=8\pi GT_{\mu\nu}, \label{2}
\end{equation} 
where $R_{\mu\nu}$, $G$ and $T_{\mu\nu}$ are the Ricci tensor, Newtonian gravitational constant and energy-momentum tensor, respectively. Within the framework of GR, a spatially flat, homogeneous and isotropic universe can be characterized by the Friedmann-Robertson-Walker spacetime
\begin{equation}
\mathrm{d}s^2=-\mathrm{d}t^2+a^2(t)\left[\frac{\mathrm{d}r^2}{1-\mathrm{K}r^2}+r^2\mathrm{d}\theta^2+r^2\mathrm{sin}^2\theta \mathrm{d}\phi^2\right],      \label{3}
\end{equation}
where $\mathrm{K}$ and $a(t)$ denote the Gaussian curvature of spacetime and the scale factor at cosmic time $t$, respectively. Substituting Eq.(\ref{3}) into Eq.(\ref{2}), we have the Friedmann equations governing the background evolution of the universe  
\begin{equation}
H^2=\frac{\sum\limits_{i}\rho_i}{3},     \label{4}
\end{equation}   
\begin{equation}
\frac{\ddot{a}}{a}=-\frac{\sum\limits_{i}(\rho_i+3p_i)}{6},     \label{5}
\end{equation}   

where $H\equiv\dot{a}/a$ is the Hubble parameter, dot is the derivative with respect to $t$, and $\rho_i$ and $p_i$ are energy densities and pressures of different matter components encompassing radiation, baryons, DM and DE. Since concentrating on the late universe, we neglect the radiation contribution to the cosmic energy budget. Furthermore, combining Eq.(\ref{4}) with Eq.(\ref{5}), we obtain the energy conservation equation as 
\begin{equation}
\dot{\rho_i}+3\frac{\dot{a}}{a}(\rho_i+p_i)=0.     \label{6}
\end{equation} 
Note that this equation can also be derived from $\Delta_\mu T^{\mu\nu}=0$. Inserting $p_i=\omega_i\rho_i$, where $\omega_i$ denotes an EoS for each matter component, one can have the corresponding Hubble parameter. Finally, we obtain the Hubble parameter of $\Lambda$CDM as
\begin{equation}
H(z)=H_0\sqrt{\Omega_{m}(1+z)^3+1-\Omega_{m}}.   \label{7}
\end{equation}
Here $H_0$ and $\Omega_{m}$ are two parameters to be confronted with data.

\begin{figure}
	\centering
	\includegraphics[scale=0.85]{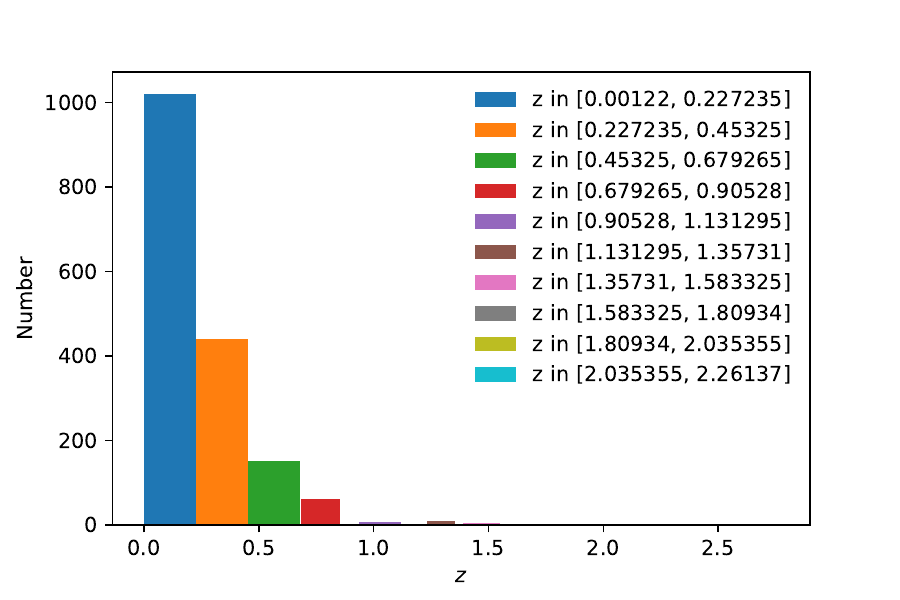}
	\caption{The redshift distribution of Pantheon+ SNe Ia for the binning method of equal redshift interval.}\label{f1}
\end{figure}

\begin{figure}
	\centering
	\includegraphics[scale=0.4]{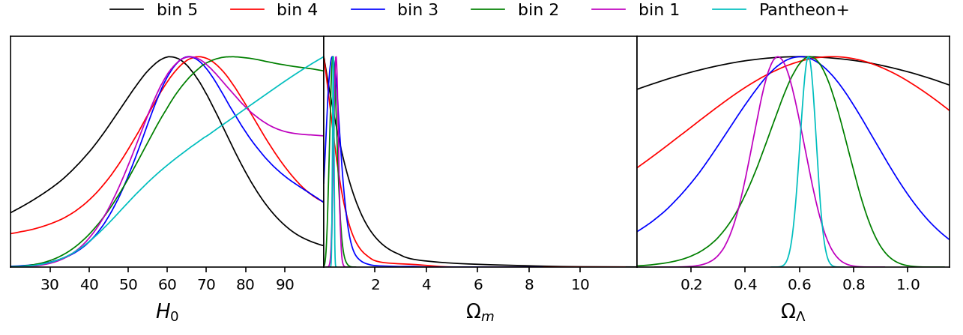}
	\caption{For the binning method of equal redshift interval, the normalized 1-dimensional posterior distributions of $H_0$, $\Omega_{m}$ and $\Omega_\Lambda$ in the $\Lambda$CDM model from the Pantheon+ sample and 5 binned subsamples without the Cepheid host distance calibration are shown, respectively.}\label{f2}
\end{figure}

\begin{figure}
	\centering
	\includegraphics[scale=0.4]{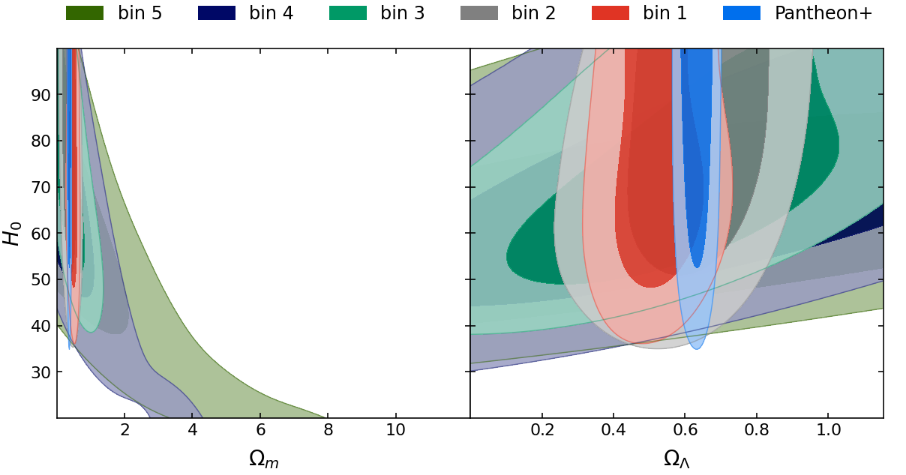}
	\caption{For the binning method of equal redshift interval, the $1\,\sigma$ and $2\,\sigma$ constraints on the $\Lambda$CDM model from the Pantheon+ sample and 5 binned subsamples without the Cepheid host distance calibration are shown, respectively.}\label{f3}
\end{figure}

\begin{figure}
	\centering
	\includegraphics[scale=0.4]{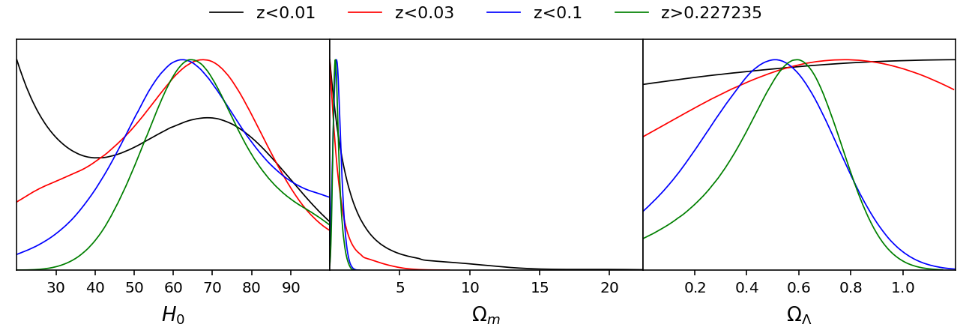}
	\caption{For the binning method of equal redshift interval, the normalized 1-dimensional posterior distributions of $H_0$, $\Omega_{m}$ and $\Omega_\Lambda$ in the $\Lambda$CDM model from 3 low-redshift and 1 high-redshift subsamples without the Cepheid host distance calibration are shown, respectively.}\label{f4}
\end{figure}

\begin{figure}
	\centering
	\includegraphics[scale=0.4]{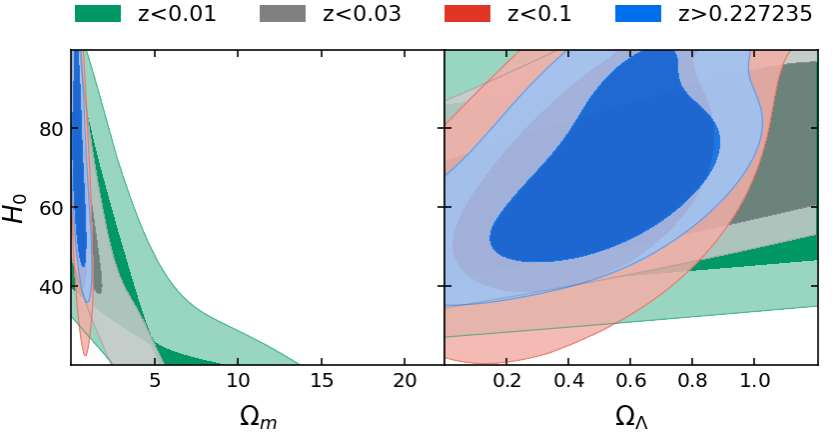}
	\caption{For the binning method of equal redshift interval, the $1\,\sigma$ and $2\,\sigma$ constraints on the $\Lambda$CDM model from 3 low-redshift and 1 high-redshift subsamples without the Cepheid host distance calibration are shown, respectively.}\label{f5}
\end{figure}

\begin{figure}
	\centering
	\includegraphics[scale=0.38]{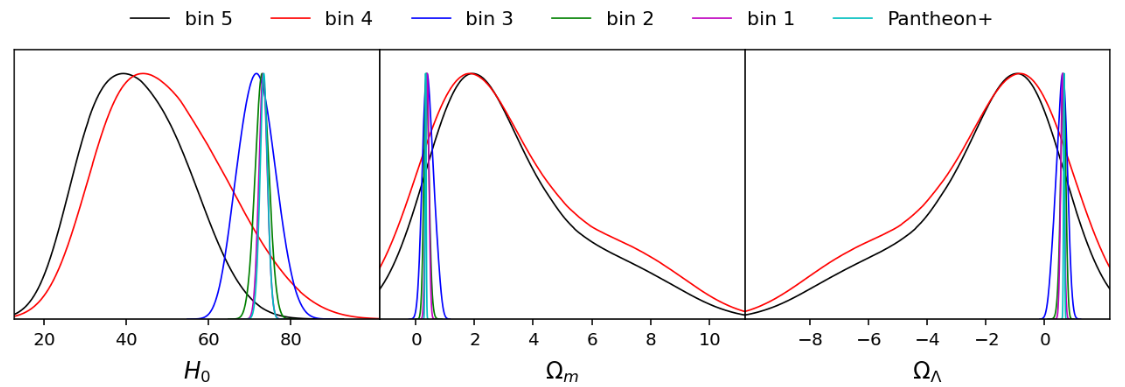}
	\caption{For the binning method of equal redshift interval, the normalized 1-dimensional posterior distributions of $H_0$, $\Omega_{m}$ and $\Omega_\Lambda$ in the $\Lambda$CDM model from the Pantheon+ sample and 5 binned subsamples with the Cepheid host distance calibration, respectively.}\label{f6}
\end{figure}

\begin{figure}
	\centering
	\includegraphics[scale=0.35]{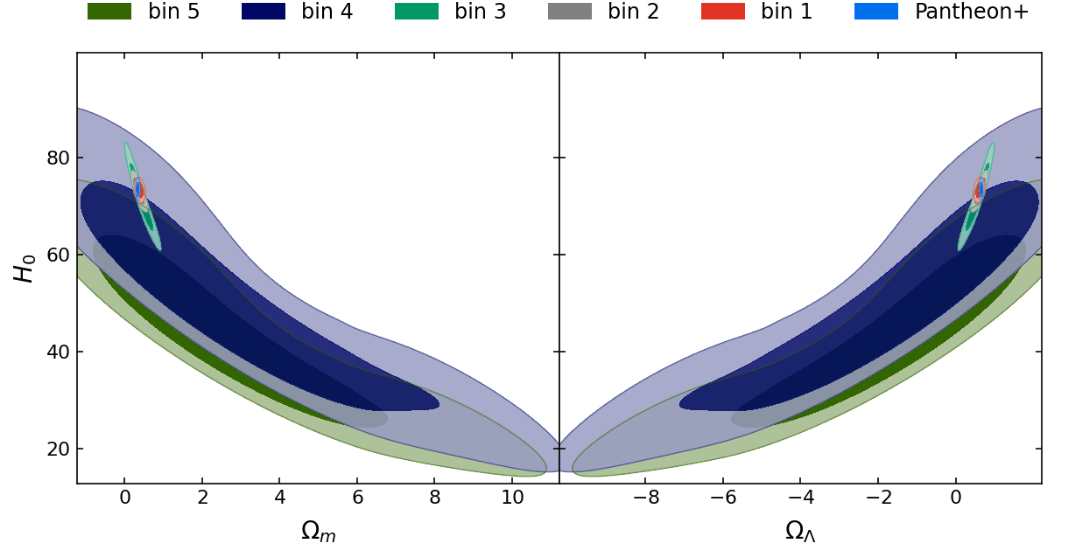}
	\caption{For the binning method of equal redshift interval, the $1\,\sigma$ and $2\,\sigma$ constraints on the $\Lambda$CDM model from the Pantheon+ sample and 5 binned subsamples with the Cepheid host distance calibration, respectively.}\label{f7}
\end{figure}

%\begin{figure}
%	\centering
%	\includegraphics[scale=0.38]{PP_5_bins_1D_calibration_adjust_range.png}
%	\caption{The normalized 1-dimensional posterior distributions of $H_0$, $\Omega_{m}$ and $\Omega_\Lambda$ in the $\Lambda$CDM model from the Pantheon+ sample and 5 binned subsamples, respectively.}\label{f8}
%\end{figure}

\begin{figure}
	\centering
	\includegraphics[scale=0.35]{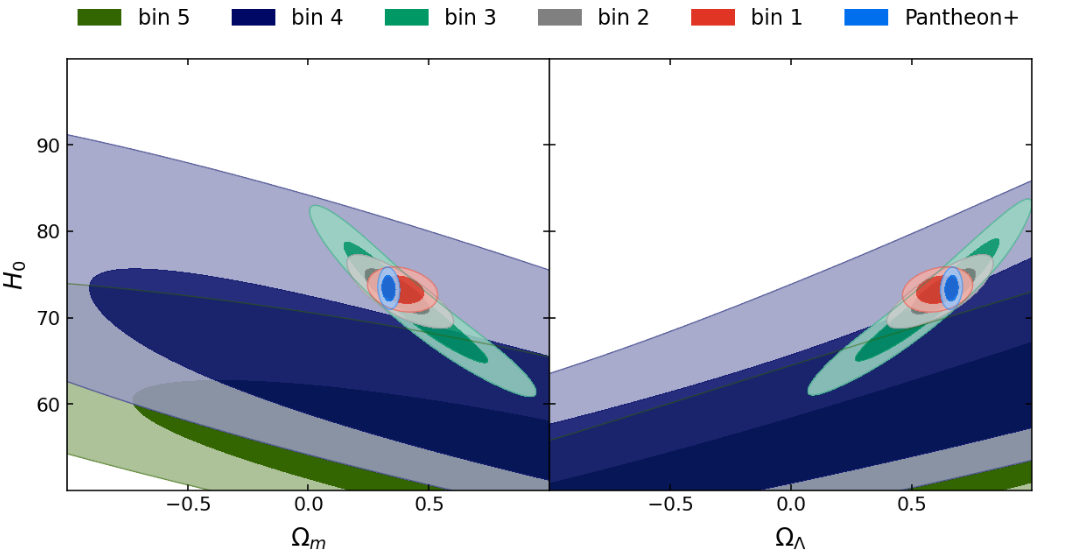}
	\caption{For the binning method of equal redshift interval, the $1\,\sigma$ and $2\,\sigma$ constraints on the $\Lambda$CDM model from the Pantheon+ sample and 5 binned subsamples with the Cepheid host distance calibration, respectively.}\label{f8}
\end{figure}

\begin{figure}
	\centering
	\includegraphics[scale=0.38]{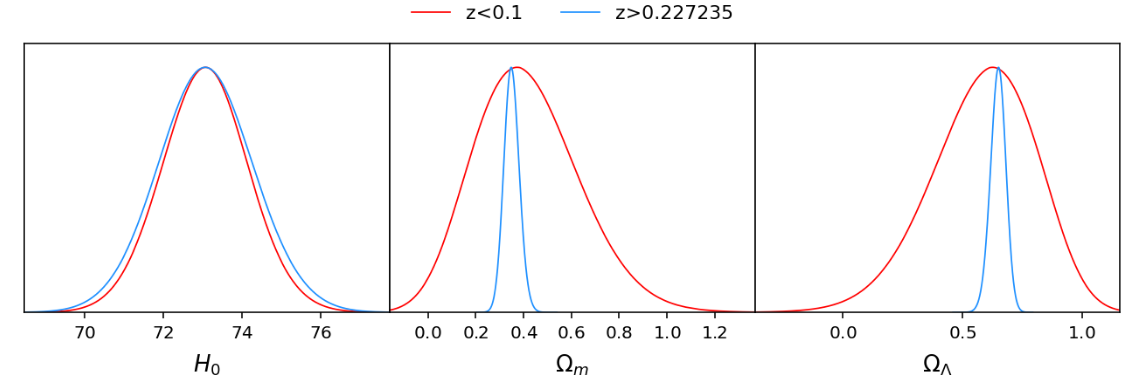}
	\caption{For the binning method of equal redshift interval, the normalized 1-dimensional posterior distributions of $H_0$, $\Omega_{m}$ and $\Omega_\Lambda$ in the $\Lambda$CDM model from 1 low-redshift and 1 high-redshift subsamples with the Cepheid host distance calibration, respectively.}\label{f9}
\end{figure}

\begin{figure}
	\centering
	\includegraphics[scale=0.35]{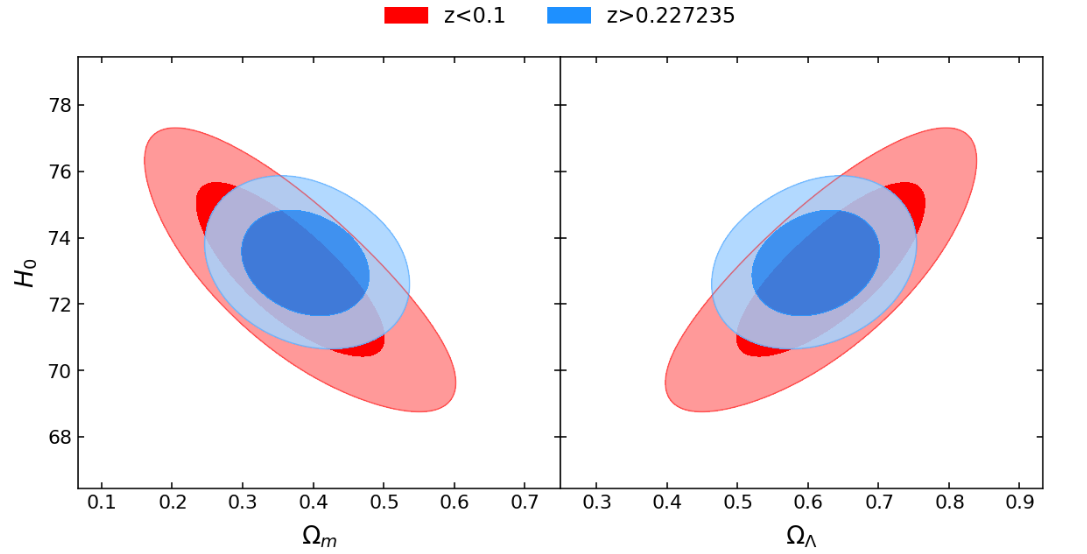}
	\caption{ For the binning method of equal redshift interval, the $1\,\sigma$ and $2\,\sigma$ constraints on the $\Lambda$CDM model from 1 low-redshift and 1 high-redshift subsamples with the Cepheid host distance calibration, respectively.}\label{f10}
\end{figure}

\begin{figure}
	\centering
	\includegraphics[scale=0.55]{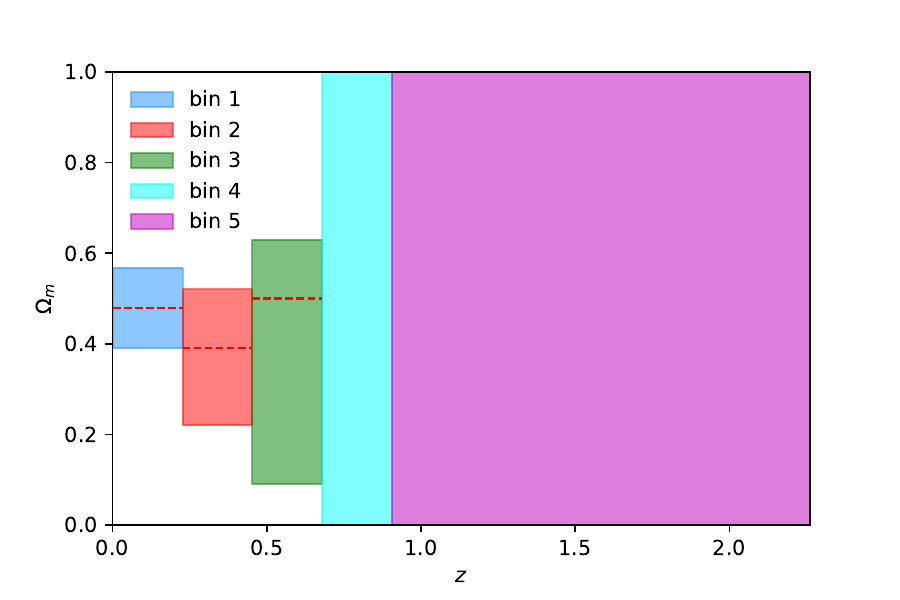}
	\includegraphics[scale=0.55]{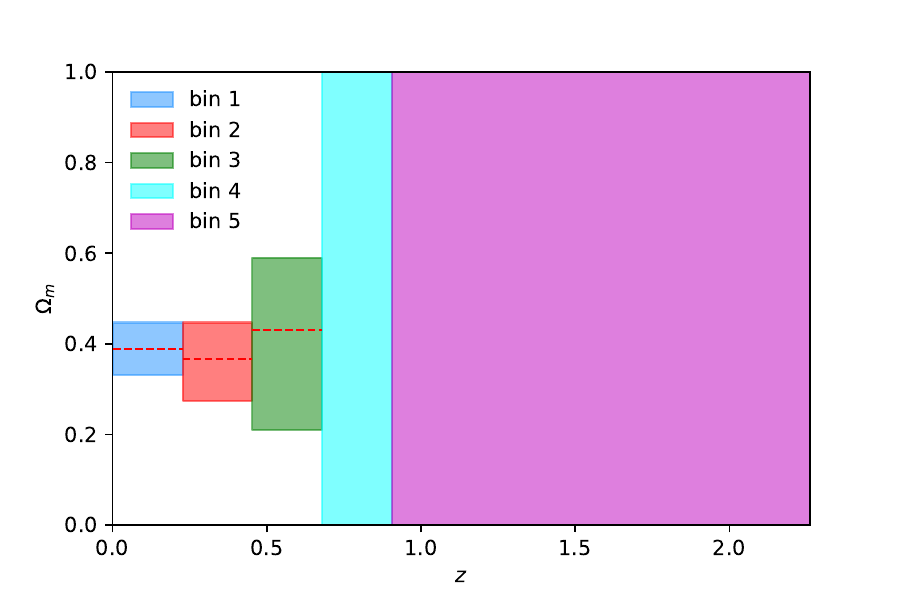}
	\caption{For the binning method of equal redshift interval, the constrained $\Omega_m$ values in 5 different redshift bins without ({\it left}) and with ({\it right})  the Cepheid host distance calibration. The dashed red lines and the shaded bands are the best fits and $1\,\sigma$ uncertainties of $\Omega_m$ in each bin, respectively.}\label{f11}
\end{figure}

\begin{figure}
	\centering
	\includegraphics[scale=0.38]{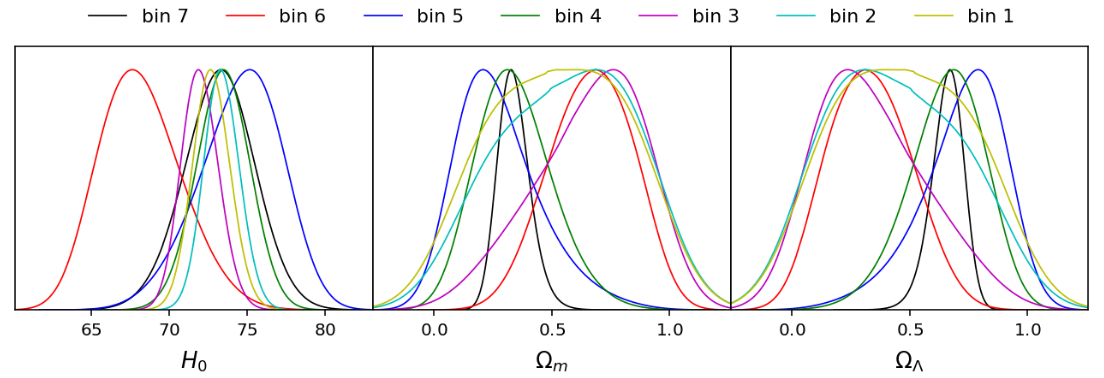}
	\caption{For the binning method of equal supernovae number, the normalized 1-dimensional posterior distributions of $H_0$, $\Omega_{m}$ and $\Omega_\Lambda$ in the $\Lambda$CDM model from the Pantheon+ sample and 7 binned subsamples with the Cepheid host distance calibration, respectively.}\label{a1}
\end{figure}

\begin{figure}
	\centering
	\includegraphics[scale=0.35]{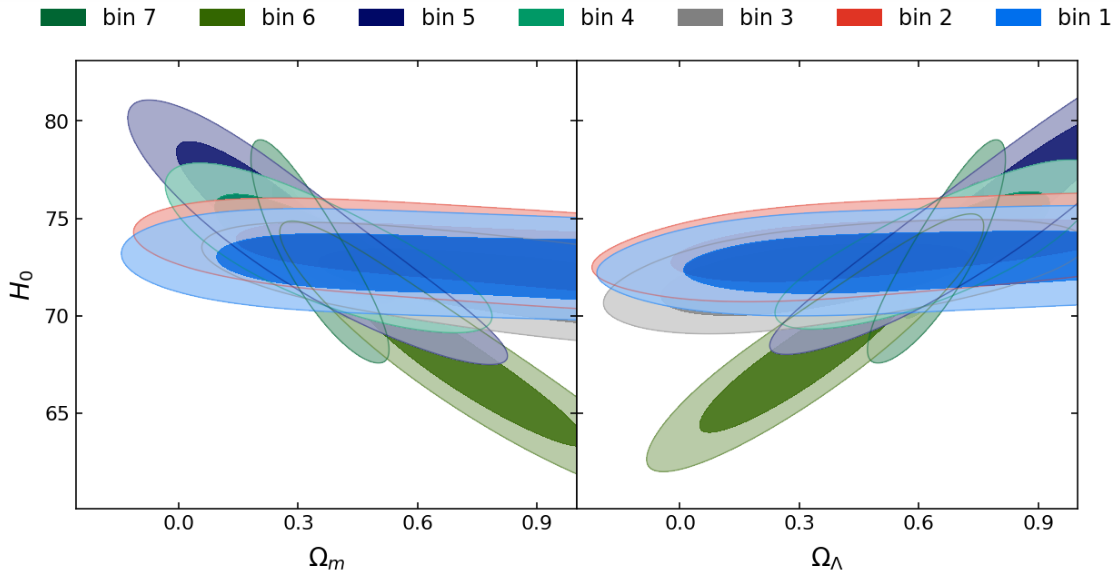}
	\caption{For the binning method of equal supernovae number, the $1\,\sigma$ and $2\,\sigma$ constraints on the $\Lambda$CDM model from the Pantheon+ sample and 7 binned subsamples with the Cepheid host distance calibration, respectively.}\label{a2}
\end{figure}

\begin{figure}
	\centering
	\includegraphics[scale=0.43]{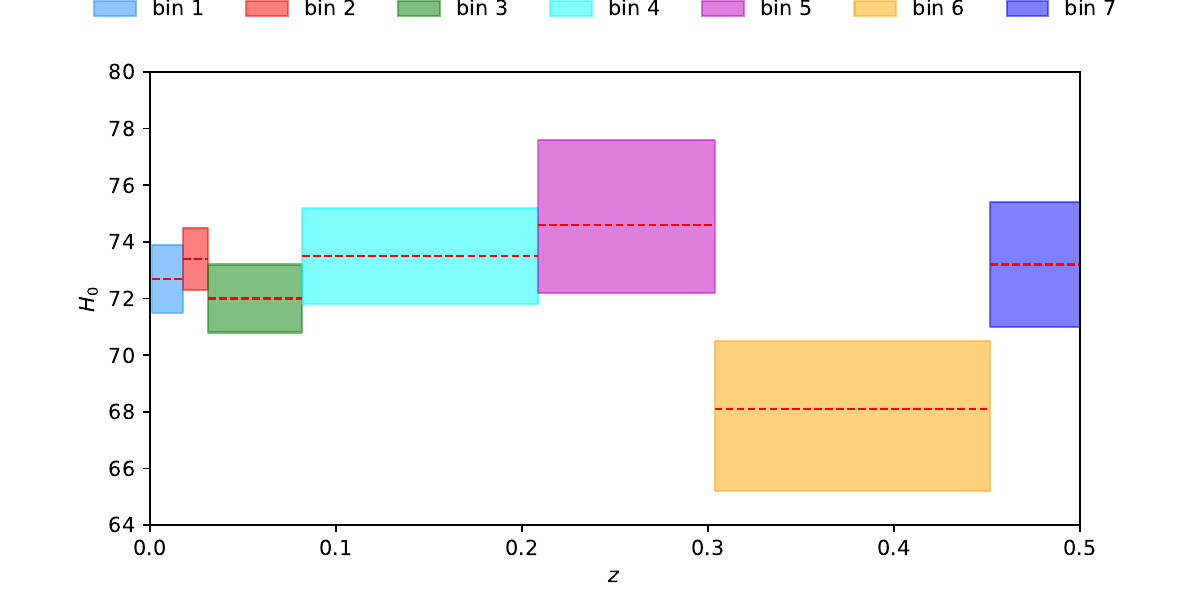}
	\includegraphics[scale=0.43]{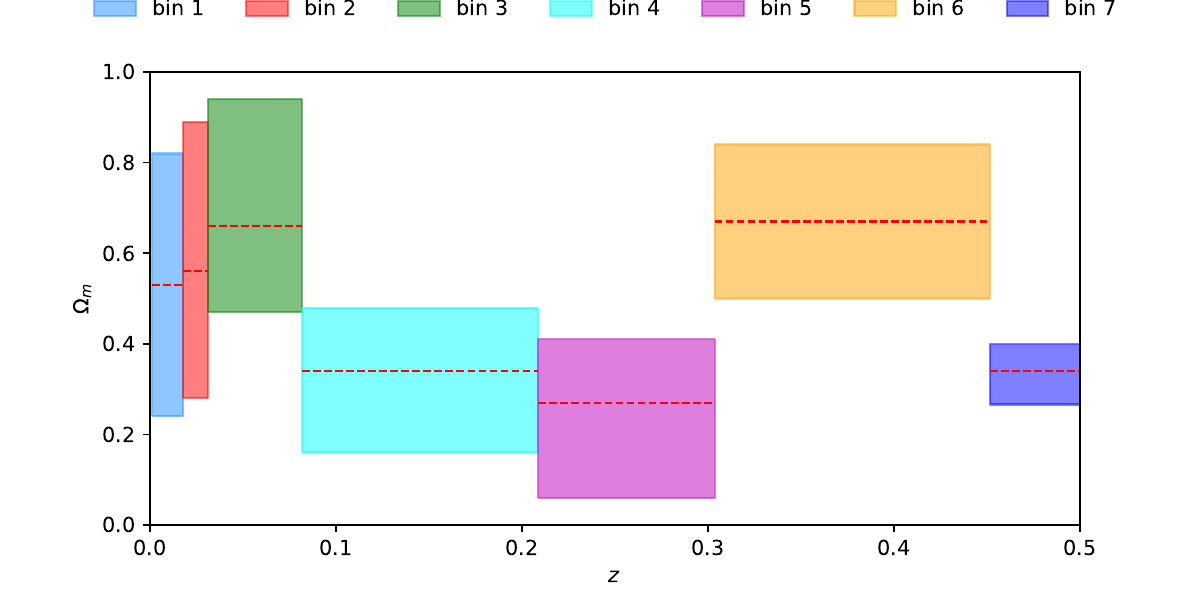}
    \caption{For the binning method of equal supernovae number, the constrained $H_0$ ({\it left}) and $\Omega_m$ ({\it right}) values in 7 different redshift bins with the Cepheid host distance calibration are shown. The dashed red lines and the shaded bands are the best fits and $1\,\sigma$ uncertainties of $H_0$ or $\Omega_m$ in each bin, respectively.}\label{a3}
\end{figure}

\section{Pantheon+ data and methodology}
The observations of SNe Ia provide a powerful tool to probe the background dynamics of the universe, particularly, the Hubble parameter and EoS of DE. As is well known, the absolute magnitudes of all SNe Ia are considered to be the same, since all SNe Ia almost explode at the same mass ($M\simeq-19.3\pm0.3$). Based on this concern, SNe Ia can be regarded as the standard candles in theory. In this study, we shall employ the latest and largest SNe Ia sample Pantheon+ to date, which consists of 1701 light curves of 1550 spectroscopically confirmed SNe Ia across 18 different surveys \cite{Brout:2022vxf}. Pantheon+ has a significant increase relative to Pantheon at low redshifts and covers the redshift range $z\in[0.00122, \, 2.26137]$. In what follows, we will consider two binning methods to analyze the Pantheon+ sample, i.e., equal redshift interval and equal supernovae number for each bin.
For the case of equal redshift interval, the distribution of Pantheon+ SNe Ia over redshift is shown in Fig.\ref{f1} and we also give the corresponding redshift range of each bin. It is easy to find that there are 4 main bins where the number of SNe Ia is larger than 50 and there are 18 SNe Ia in the last six bins. It is worth noting that the redshifts in Fig.\ref{f1} have six digits after the decimal point is because we divide evenly the redshift range $z\in[0.00122, \, 2.26137]$ into 10 bins.
For the case of equal supernovae number, we divide the whole redshift range into 7 bins, where each bin includes 243 SNe Ia data points.

The theoretical distance modulus for a SNe Ia is defined as 
\begin{equation}
\mu_{\mathrm{th}}(z)=5\mathrm{log}_{10}D_\mathrm{L}(z)+25,   \label{8}
\end{equation}
where the luminosity distance reads as
\begin{equation}
D_\mathrm{L}(z)=\frac{c\,(1+z)}{H_0}\int_{0}^{z}\frac{\mathrm{d}z'}{E(z')},   \label{9}
\end{equation} 
where $c$ is the speed of light. Then, the $\chi^2$ can be easily expressed as 
\begin{equation}
\chi^2 = (\mathbf{\mu}_{\mathrm{th}}-\mathbf{\mu}_{\mathrm{obs}})\mathcal{C}^{-1}(\mathbf{\mu}_{\mathrm{th}}-\mathbf{\mu}_{\mathrm{obs}})^{\mathrm{T}}, \label{10}
\end{equation}
where the superscript $\mathrm{T}$ represents the transpose of a vector or a matrix, $\mathcal{C}$ is the covariance matrix, and $\mathbf{\mu}_{\mathrm{obs}}=m_b-M$ denotes the observed distance modulus of each SNe Ia. $m_b$ and $M$ are the apparent magnitude and the absolute magnitude, respectively.   

In light of this distribution displayed in Fig.\ref{f1}, we will divide the Pantheon+ sample into five bins and perform a tomographic analysis. Note that the redshift range of the fifth bin is $z\in[0.90528, \, 2.26137]$. Subsequently, to investigate the constraining power of the low-redshift and high-redshift SNe Ia in the Pantheon+ sample, we constrain $\Lambda$CDM with data lying in $z<0.01$, $z<0.03$, $z<0.1$ and $z>0.227235$, respectively. Moreover, we also carry out a global fitting to explore the full six parameter space of $\Lambda$CDM. Besides Pantheon+ (hereafter ``P''), the extra datasets used are listed below:

$\bullet$ CMB: Observations from the Planck satellite  have measured the cosmic matter components, topology and large scale structure of the universe. We use the Planck-2018 CMB temperature and polarization data including the likelihoods of temperature at $30\leqslant \ell\leqslant 2500$ and the low-$\ell$ temperature and polarization likelihoods at $2\leqslant \ell\leqslant 29$, i.e., TTTEEE$+$lowE, and Planck-2018 CMB lensing data \cite{Planck:2018vyg}. This dataset is denoted as ``C''.

$\bullet$ BAO: BAO as a standard ruler to measure the background dynamics of the universe is hardly affected by uncertainties in the nonlinear evolution of matter density field and other systematic errors. To break the parameter degeneracy from other observations, we employ 4 BAO data points: the 6dFGS sample at effective redshift $z_{eff}=0.106$ \cite{Beutler:2011hx}, the SDSS-MGS one at $z_{eff}=0.15$ \cite{Ross:2014qpa}, and the BOSS DR12 dataset at three effective redshifts $z_{eff}=$ 0.38, 0.51 and 0.61 \cite{BOSS:2016wmc}. We refer to this dataset as ``B''.

$\bullet$ Cosmic chronometers (CC): We take the direct observations of cosmic expansion rate from CC as a complementary probe, which has no cosmology dependence. This dataset is obtained by using the most massive and passively evolving galaxies based on the ``galaxy differential age'' method. We adopt 31 CC data points with systematic uncertainties to help constrain $\Lambda$CDM \cite{Moresco:2020fbm}. This dataset is identified as ``H''. 

We also include the following three 2-point correlation functions measured from the Dark Energy Survey Year 1 (DES Y1) \cite{DES:2017myr,DES:2018ufa}:

$\bullet$ Galaxy clustering: The homogeneity of matter density field in the universe can be traced via galaxies
distribution. The overabundance of pairs at angular separation $\theta$ in a random distribution, $\omega(\theta)$, is one of the most convenient approaches to measure galaxy clustering. It quantifies the scale dependence and strength of galaxy clustering,
and consequently affects the matter clustering \cite{DES:2017qwj}.

$\bullet$ Cosmic shear: The 2-point statistics characterizing the shapes of galaxies are very complicated, because they are products of components of a spin-2 tensor. Therefore, it is convenient to extract information from a galaxy survey by using a pair of 2-point correlation functions $\xi_+(\theta)$ and $\xi_-(\theta)$, which denote the sum and difference of products of tangential and cross components of the shear, measured along the line connecting each pair of galaxies \cite{DES:2017hdw}.

$\bullet$ Galaxy-galaxy lensing: The characteristic distortion of source galaxy shapes is originated from mass associated with foreground lenses. This typical distortion is the mean tangential ellipticity of source galaxy shapes around lens galaxy positions for each
pair of redshift bins and also called as the tangential shear, $\gamma_t(\theta)$ \cite{DES:2017gwu}.

More details about the DES Y1 dataset can be found in \cite{DES:2017qwj,DES:2017hdw,DES:2017gwu}. Hereafter this dataset is denoted as ``W''.

\begin{table*}[!t]
	\renewcommand\arraystretch{1.5}
	\caption{For the binning method of equal redshift interval, the confidence ranges of two parameters $H_0$ and $\Omega_{m}$ in the $\Lambda$CDM model are shown for each separate bin or their combination. The symbols $\bigstar$ denote the parameters that cannot be well constrained by observations.}
	\setlength{\tabcolsep}{10mm}{
		\begin{tabular} { l |c| c |c}
			\hline
			\hline
			
			Data               &Range          & $H_0$      & $\Omega_m$                             \\
			\hline
			Pantheon+          &[0.00122, \, 2.26137]            &$>45.7$ $(2\,\sigma)$     &$0.367\pm0.030$    \\
			\hline
			bin 1          &[0.00122, \, 0.227235]            &$>46.7$ $(2\,\sigma)$     &$0.479\pm0.089$     \\
			\hline
			bin 2          &[0.227235, \, 0.45325]            &$>46.9$ $(2\,\sigma)$     &$0.39^{+0.13}_{-0.17}$     \\
			\hline
			bin 3          &[0.45325, \, 0.679265]           &$>48.2$ $(2\,\sigma)$     &$0.50^{+0.13}_{-0.41}$     \\
			\hline		    
			bin 4          &[0.679265, \, 0.90528]            & $66\pm20$    &$<2.16$ $(2\,\sigma)$     \\
			\hline		    
			bin 5          &[0.90528, \, 2.26137]            &$58\pm20$     &$<4.48$ $(2\,\sigma)$     \\
			\hline		    
			Low-z bin 1    &[0.00122, \, 0.01]            & $\bigstar$     &$<10.0$ $(2\,\sigma)$     \\
			\hline		                                
			Low-z bin 2    &[0.00122, \, 0.03]            & $\bigstar$     &$<3.52$ $(2\,\sigma)$     \\
			\hline
			Low-z bin 3    &[0.00122, \, 0.1]            & $>35.0$ $(2\,\sigma)$    &$0.57^{+0.22}_{-0.35}$     \\
			\hline
			High-z bin 1   &[0.227235, \, 2.26137]            &$67^{+10}_{-20}$     &$0.50^{+0.16}_{-0.30}$     \\
			\hline
			\hline
		\end{tabular}
		\label{t1}}
\end{table*}

\begin{table*}[!t]
	\renewcommand\arraystretch{1.5}
	\caption{For the binning method of equal redshift interval, the confidence ranges of two parameters $H_0$ and $\Omega_{m}$ in the $\Lambda$CDM model are shown for each separate bin or their combination with the Cepheid host distance calibration.}
	\setlength{\tabcolsep}{10mm}{
		\begin{tabular} { l |c| c |c}
			\hline
			\hline
			
			Data               &Range          & $H_0$      & $\Omega_m$                             \\
			\hline
			Pantheon+          &[0.00122, \, 2.26137]            &$73.4\pm1.1$     &$0.337\pm0.018$    \\
			\hline
			bin 1          &[0.00122, \, 0.227235]            &$73.2\pm1.1$     &$0.389\pm0.058$     \\
			\hline
			bin 2          &[0.227235, \, 0.45325]            &$73.0\pm1.7$     &$0.367^{+0.080}_{-0.093}$     \\
			\hline
			bin 3          &[0.45325, \, 0.679265]           &$71.6\pm4.4$     &$0.43^{+0.16}_{-0.22}$     \\
			\hline		    
			bin 4          &[0.679265, \, 0.90528]            &$49^{+10}_{-20}$    &$3.1^{+1.8}_{-2.4}$     \\
			\hline		    
			bin 5          &[0.90528, \, 2.26137]            &$42\pm10$     &$3.1^{+1.6}_{-3.1}$     \\	    
			\hline
			Low-z bin 3    &[0.00122, \, 0.1]            &$73.1\pm1.1$    &$0.41^{+0.19}_{-0.24}$     \\
			\hline
			High-z bin 1   &[0.227235, \, 2.26137]            &$73.1\pm1.2$     &$0.351\pm0.032$     \\
			\hline
			\hline
		\end{tabular}
		\label{t2}}
\end{table*}

\begin{figure}
	\centering
	\includegraphics[scale=0.45]{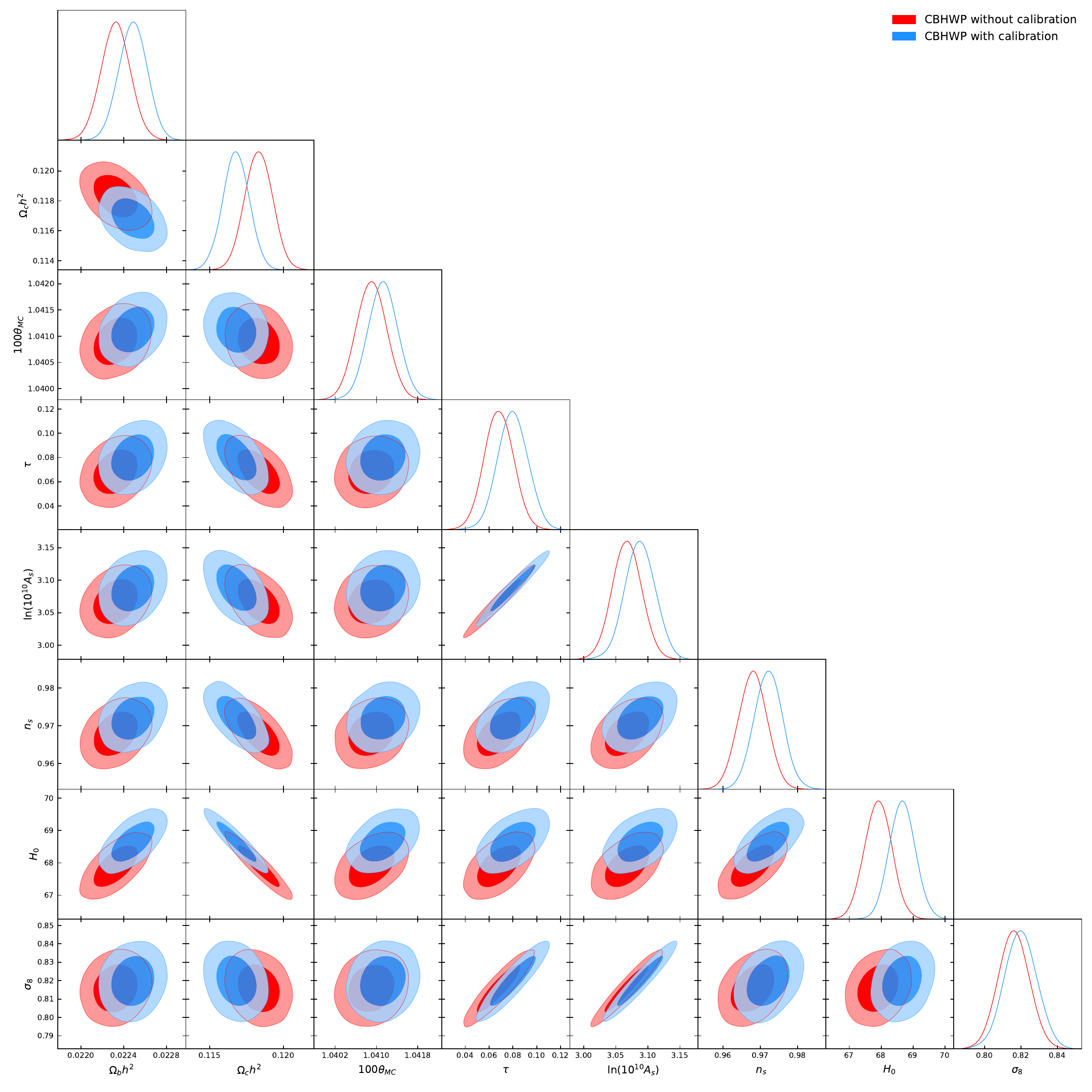}
	\caption{The $1\,\sigma$ and $2\,\sigma$ constraints on the $\Lambda$CDM model from the combined datasets CBHWP with (blue) and without (red) with the Cepheid host distance calibration. Here we also display the correlations between six basic parameters and two derived parameters $H_0$ and $\sigma_8$.}\label{f12}
\end{figure}

\begin{table*}[!t]
	\renewcommand\arraystretch{1.5}
	\caption{For the binning method of equal supernovae number, the confidence ranges of two parameters $H_0$ and $\Omega_{m}$ in the $\Lambda$CDM model are shown for each separate bin with the Cepheid host distance calibration.}
	\setlength{\tabcolsep}{10mm}{
		\begin{tabular} { l |c| c |c}
			\hline
			\hline
			
			Data               &Range          & $H_0$      & $\Omega_m$                             \\
			\hline
			bin 1          &[0.00122, \, 0.01778]            &$72.7\pm1.2$     &$0.53\pm0.29$     \\
			\hline
			bin 2          &[0.01778, \, 0.03110]            &$73.4\pm1.1$     &$0.56^{+0.33}_{-0.28}$     \\
			\hline
			bin 3          &[0.03110, \, 0.08167]           &$72.0\pm1.2$     &$0.66^{+0.28}_{-0.19}$     \\
			\hline		    
			bin 4          &[0.08167, \, 0.20879]            &$73.5\pm1.7$    &$0.34^{+0.14}_{-0.18}$     \\
			\hline		    
			bin 5          &[0.20879, \, 0.30352]            &$74.6^{+3.0}_{-2.4}$     &$0.27^{+0.14}_{-0.21}$     \\	    
			\hline
			bin 6    &[0.30352, \, 0.45162]            &$68.1^{+2.4}_{-2.9}$    &$0.67\pm0.17$     \\
			\hline
			bin 7   &[0.45162, \, 2.26137]            &$73.2\pm2.2$     &$0.340^{+0.059}_{-0.074}$     \\
			\hline
			\hline
		\end{tabular}
		\label{t3}}
\end{table*}

\begin{table*}[!t]
	\renewcommand\arraystretch{1.5}
	\caption{The $1\,\sigma$ confidence ranges of parameters in the $\Lambda$CDM model are shown by using the combined datasets CBHWP with and without the Cepheid host distance calibration.}
	\setlength{\tabcolsep}{5mm}{
		\begin{tabular} { l |c| c }
			\hline
			\hline
			
			Parameter               &Without calibration     & With calibration    \\
			\hline
			$\Omega_bh^2$    & $0.02232\pm0.00014$          & $0.02249\pm0.00013$                \\ 
			\hline  
			$\Omega_ch^2$    &  $0.11832\pm0.00094$         &  $0.11679\pm0.00089$               \\                          
			\hline
			$100\theta_{MC}$  &  $1.04090\pm0.00029$         & $1.04113\pm0.00029$                \\
			\hline
			$\mathrm{ln}(10^{10}A_s)$  & $3.063\pm0.023$          &  $3.088\pm0.024$               \\
			\hline
			$n_s$         & $0.9680\pm0.0038$          & $0.9722\pm0.0038$                \\
			\hline
			$\tau$     & $0.068\pm0.012$          &  $0.080\pm0.013$               \\
			\hline

			$H_0$      &  $67.90\pm0.42$          & $68.66\pm0.41$                \\
			\hline
			$\sigma_8$   & $0.8160\pm0.0085$           & $0.8197\pm0.0090$                 \\    
			\hline
			\hline
		\end{tabular}
		\label{t4}}
\end{table*}

In order to implement a detailed analysis of Pantheon+ sample, for the case of equal redshift interval, a complete strategy is constraining $\Lambda$CDM in two cases: (i) without the Cepheid host distance calibration; (ii) with the Cepheid host distance calibration. As a consequence, when not considering the calibration, we constrain $\Lambda$CDM with each separate bin or their combination (i.e., full Pantheon+ sample). However, for simplicity, we just consider the case with the Cepheid host distance calibration for the binning method of equal supernovae number. Specifically,
we take the Bayesian analysis to derive the posterior distributions of free parameters. The priors of free parameters are $H_0\in[20,\,100]$, $\Omega_{m}\in[0,15]$ and $M\in[-25, -15]$. To implement a global constraint on six-parameter space using the data combination of C, B, H, W and P (hereafter CBHWP), we use the code \texttt{CosmoMC} \cite{Lewis:2013hha} and the corresponding priors we use are $\Omega_bh^2 \in [0.005, 0.1]$, $\Omega_ch^2 \in [0.001, 0.99]$, $100\theta_{MC} \in [0.5, 10]$,  $\mathrm{ln}(10^{10}A_s) \in [2, 4]$, $n_s \in [0.8, 1.2]$, $\tau \in [0.01, 0.8]$, where $\Omega_bh^2$ and $\Omega_ch^2$
denote the present-day baryon and CDM densities, $\theta_{MC}$ is the ratio between angular diameter
distance and sound horizon at the redshift of last scattering, $\tau$ is the optical depth due to the reionization, and $A_s$ and $n_s$ are the amplitude and spectral index of primordial scalar power spectrum. We adopt the Markov Chain Monte Carlo (MCMC) method to sample the parameter space and use the public package \texttt{Getdist} to analyze the MCMC chains \cite{Lewis:2019xzd}. Since SNe Ia can not give any constraint on $H_0$, we shall consider the SH0ES distance calibration. Keeping the above priors for the case without the SH0ES calibration unchanged except $\Omega_{m}\in[-5,15]$, we replace the original 77 data points with 77 Cepheid host distance modulus \cite{Brout:2022vxf} and redo the constraints on $\Lambda$CDM.  

\section{Numerical results}
For the binning method of equal redshift interval, the numerical results of tomographic analysis of the Pantheon+ sample without the SH0ES distance calibration are presented in Figs.\ref{f2}-\ref{f5} and Tab.\ref{t1}. 
For the Pantheon+ data alone, using the relatively weak priors $H_0\in[20,\,100]$, $\Omega_{m}\in[0,15]$ and $M\in[-25, -15]$, we obtain the $2\,\sigma$ lower bound on the Hubble constant $H_0>45.7$ km s$^{-1}$ Mpc$^{-1}$ and the matter fraction $\Omega_m=0.367\pm0.030$. This means that the enhanced SNe Ia sample can not give a better constraint on $H_0$ when considering the weak priors of $H_0$ and $M$, but can provide a 8\% precision determination of matter density ratio. However, this $\Omega_m$ value is in a $1.7\,\sigma$ tension with that from the Planck-2018 measurement \cite{Planck:2018vyg}. As indicated in Ref.\cite{Brout:2022vxf}, after adding the $H_0$ prior from the SH0ES collaboration \cite{Riess:2021jrx}, $\Omega_m$ will decrease to $0.338\pm0.018$. Interestingly, with increasing redshift and decreasing number of SNe Ia, first three bins all give higher $2\,\sigma$ lower bounds on $H_0$ (see Fig.\ref{2} and Tab.\ref{t1}). This does not mean that they have stronger constraining power than the full sample, and the price is larger $\Omega_m$ values with larger uncertainties. Different from the first three bins, one can find that bin 4 and bin 5 can give $1\,\sigma$ constraints on $H_0$, although errors are very large. At beyond $1\,\sigma$ level, they give different constraints on $H_0$ and $\Omega_m$ from bin 1 and the full sample. 
Due to low sample sizes, they give unphysical $2\,\sigma$ upper bounds on the matter fraction $\Omega_{m}$, which must be less than 1. Hence, we should be very cautious to the relatively tight constraint on $H_0$ from the last two bins. From Fig.\ref{f3}, we also observe that the constraining power increases with the increasing SNe Ia number. Moreover, 1021 SNe Ia in bin 1 dominate the final constraining power of the whole Pantheon+ sample due to the smallest error of $\Omega_{m}$ among five bins. This is consistent with the prediction of its dominant sample size in Pantheon+. After adding the left low-z data, both the best fitting value of $\Omega_{m}$ and its corresponding error decrease. Moreover, we derive $\Omega_\Lambda=0.633\pm0.030$ from $\Omega_m=0.367\pm0.030$, which gives the evidence of DE at the $21\,\sigma$ confidence level.      

Since the significant enhancement of low-redshift SNe Ia abundance in Pantheon+, we are very interested in the effects of low-z and hig-z data on $H_0$ and $\Omega_{m}$. From Figs.\ref{3} and \ref{4}, we find that the first low-z bins can not constrain $\Lambda$CDM well, especially gives unphysical constraints on $\Omega_{m}$. Nonetheless, the low-z bin 3 with 741 SNe Ia gives a $2\,\sigma$ constraint $H_0>35.0$ km s$^{-1}$ Mpc$^{-1}$ and $68\%$ confidence range $\Omega_{m}=0.57^{+0.22}_{-0.35}$. This means that low-redshift SNe Ia in Pantheon+ can not give good constraint on $\Lambda$CDM, and high-redshift data provides a better constraint than low-redshift observations. To verify this, we choose a subsample of $z>0.227235$ and give $1\,\sigma$ constraints $67^{+10}_{-20}$ km s$^{-1}$ Mpc$^{-1}$ and $\Omega_{m}=0.50^{+0.16}_{-0.30}$. It is easy to find that the constraint gets better and the parameter space is obviously compressed in the high-z range (see Fig.\ref{f5}).   

When considering the SH0ES calibration, the results are displayed in Figs.\ref{f6}-\ref{f11} and Tab.\ref{t2}. One can see we give strong constraints $H_0=73.4\pm1.1$ km s$^{-1}$ Mpc$^{-1}$ and $\Omega_m=0.337\pm0.018$, which gives an evidence of DE at the $18.7\,\sigma$ confidence level and is completely consistent with the result from Ref.\cite{Brout:2022vxf}. From Figs.\ref{f6} and \ref{f7}, one can easily observe that the first three bins have much stronger limitations to $H_0$ and $\Omega_m$ than they do in the case without the calibration, except the last high-redshift bins. This implies that the SH0ES calibration has a very strong limitation to the background parameter space. The inclusion of it reduces obviously the constrained values of $H_0$ and $\Omega_m$ in each bin (see Tab.\ref{t2} and Fig.\ref{f8} for a zoom-in version). Nonetheless, similar to the case without the calibration, higher redshift bins show a weaker constraining power than lower bins and the first bin still dominates the constraining power of the full sample as predicted. Interestingly, the $\Omega_m$ value can be smaller than 0 for bin 4 and bin 5, due to the lack of SNe Ia data points. Furthermore, we find that current $H_0$ tension can be well resolved to $0.96\,\sigma$ by bin 3, which gives $H_0=71.6\pm4.4$ km s$^{-1}$ Mpc$^{-1}$. We think this alleviation is mainly attributed to the inclusion of the SH0ES calibration and the large error bars from bin 3.     

For simplicity, to study the impacts of low-z and hig-z data on the background evolution, we just consider $z\in[0.00122, \, 0.1]$ and $z\in[0.227235, \, 2.26137]$ in the case with the calibration, which includes 741 and 680 SNe Ia, respectively. We find that they give close constraints on $H_0$ but different limitations to $\Omega_{m}$. Obviously, this High-z bin gives $\Omega_m=0.351\pm0.032$ that is tighter than $\Omega_m=0.389\pm0.058$ from bin 1 which contains 1201 SNe Ia. This indicates that high-z data ($z>0.227235$) have a strong constraining power than low-z SNe Ia. The constraining power of both bins are clearly presented in Figs.\ref{f9} and \ref{f10}.

In Fig.\ref{f11}, we show directly the constrained $\Omega_m$ values in different redshift bins. It is easy to see that there is no evolution of $\Omega_m$ over $z$. All the values agree with each other at the $1\,\sigma$ confidence level regardless of whether we consider the SH0ES calibration or not. 

For the binning method of equal supernovae number, our numerical results with the SH0ES distance calibration are exhibited in Figs.\ref{a1}-\ref{a3} and Tab.\ref{t3}. In Fig.\ref{a1}, we find that bin 6 and bin 5 give a lower $H_0$ and slightly larger $H_0$ values than other bins do, respectively. Bin 1 and bin 2 can not provide strong constraining power for cosmic matter density $\Omega_m$. In Fig.\ref{a2}, bins 1, 2, 3, and 6 prefer a larger $\Omega_m$ than the left bins (see also Tab.\ref{t3}) and, interestingly, these four bins can only give very weak evidences of DE at less than $2\,\sigma$ confidence level. Subsequently, we obtain $\Omega_m=0.340^{+0.059}_{-0.074}$ from bin 7, which shows the good constraining power again from high-z SNe Ia data. Furthermore, similar to the case of equal redshift interval, we also analyze the cosmic expansion rate and matter density in different redshift bins. In Fig.\ref{a3}, it is easy to see that there is a $1\,\sigma$ $H_0$ gap between bin 6 and other bins and that the same consequence occurs in the $\Omega_m$-$z$ panel. However, this can not be an evidence of evolution of these two cosmological parameters. It just gives a possible clue of evolving parameters through the late-time history of the universe. There is still no evolution of $H_0$ and $\Omega_m$ at $2\,\sigma$ confidence level. Moreover, compared to the case of equal redshift interval, equal bin size can give more stable constraints across different redshifts bins, since it does not lead to obviously small subsample size.

In light of this large SNe Ia sample, we are also very interested in giving the most stringent constraint on $\Lambda$CDM. Using the combined datasets CBHWP with and without the calibration, we give the constraining results of parameters in Fig.\ref{f12} and Tab.\ref{t4}. For the case with the calibration, we provide today's cosmic expansion rate $H_0=67.90\pm0.42$ km s$^{-1}$ Mpc$^{-1}$ and the matter fluctuation amplitude $\sigma_8=0.8160\pm0.0085$, which is very consistent with the Planck-2018 results at the $1\,\sigma$ confidence level. Interestingly, when considering the calibration, $H_0=68.66\pm0.42$ km s$^{-1}$ Mpc$^{-1}$, which is in a $\sim 2\,\sigma$ tension with the Planck-2018 measurement $H_0=67.36\pm0.54$ km s$^{-1}$ Mpc$^{-1}$. One can observe that the inclusion of the SH0ES calibration leads to a global shift of the parameter space.

\section{Discussions and conclusions}
The recently released Pantheon+ SNe Ia sample can provide stronger constraining power on the background evolution of the late universe than the original Pantheon sample. We implement a tomographic analysis to explore the origin of the enhanced constraining power and internal correlations of datasets in different redshifts.  

For the binning method of equal redshift interval, in the case without the SH0ES calibration, using the weak priors on $H_0$, $\Omega_{m}$ and $M$, we give the $2\,\sigma$ lower bound on the present-day cosmic expansion rate $H_0>45.7$ km s$^{-1}$ Mpc$^{-1}$ and the matter density ratio $\Omega_m=0.367\pm0.030$, which gives the evidence of DE at the $21\,\sigma$ confidence level but is in a $1.7\,\sigma$ tension with that from the Planck-2018 measurement.
After dividing the full sample to 10 bins, we find that the first bin in the redshift range $z\in[0.00122, \, 0.227235]$ dominates the constraining power of the whole sample (see Tab.\ref{t1}). However, it produces a relative high matter fraction $0.479\pm0.089$ even though giving a small error relative to other bins. This means that the effect of left 9 bins is actually pulling $\Omega_m$ to a lower value and reducing the statistical uncertainties. We also give constraining results from other bins and find that $\Lambda$CDM can not be well constrained with them.

In the case with the SH0ES calibration, we obtain tight constraints $H_0=73.4\pm1.1$ km s$^{-1}$ Mpc$^{-1}$ and $\Omega_m=0.337\pm0.018$, which produces an evidence of DE at the $18.7\,\sigma$ confidence level, is completely consistent with the result from Ref.\cite{Brout:2022vxf}, and is well compatible with the Planck-2018 measurement at $\sim 1\,\sigma$ confidence level. The first three bins reduce the parameter space of background dynamics of the universe and give very strong constraints on $H_0$ and $\Omega_m$. Similar to the case without the calibration, the first bin still dominates the constraining power of the whole sample and higher redshift bins exhibit a weaker constraining power than lower ones. It is interesting that bin 3 can alleviate the $H_0$ tension to $0.96\,\sigma$. Although this may be due to the large data uncertainties from bin 3, it give a new clue towards the final solution of the $H_0$ tension, i.e., searching for the possible solution in $z\in [0.45325, \, 0.679265]$.

By studying the impacts of low-z and hig-z data on the background evolution, we may conclude that the Pantheon+ sample provides the constraining power of $H_0$ and $\Omega_m$ using SNe Ia data lying in $z<0.227235$ and $z>0.227235$, respectively. 

For the binning method of equal supernovae number with the SH0ES calibration, we obtain the tight constraint $H_0=73.4\pm1.1$ km s$^{-1}$ Mpc$^{-1}$ and but a slightly large $\Omega_m=0.56^{+0.33}_{-0.28}$ in the second bin. This binning method can give more stable constraints across different redshifts bins, because it does not lead to small enough subsample size.

Furthermore, we observe that Pantheon+ excludes the evolution of $H_0$ and $\Omega_{m}$ at $2\,\sigma$ confidence level regardless of whether we consider the SH0ES calibration or not. We also give the most stringent constraint on $\Lambda$CDM using the combined dataset CBHWP and find that the inclusion of the SH0ES calibration leads to a global shift of the parameter space. This suggests that the Cepheid host distance calibration, which affects largely the measurement of $H_0$ value, will obviously affect our knowledge about the evolution of the universe.

\section{Acknowledgments}
DW is supported by the National Science Foundation of China under Grants No.11988101 and No.11851301.


\begin{thebibliography}{99}
%\cite{SupernovaSearchTeam:1998fmf}
\bibitem{SupernovaSearchTeam:1998fmf}
A.~G.~Riess \textit{et al.} [Supernova Search Team],
``Observational evidence from supernovae for an accelerating universe and a cosmological constant,''
Astron. J. \textbf{116}, 1009-1038 (1998).
	
%\cite{SupernovaCosmologyProject:1998vns}
\bibitem{SupernovaCosmologyProject:1998vns}
S.~Perlmutter \textit{et al.} [Supernova Cosmology Project],
``Measurements of $\Omega$ and $\Lambda$ from 42 high redshift supernovae,''
Astrophys. J. \textbf{517}, 565-586 (1999).

%\cite{DiValentino:2020vhf}
\bibitem{DiValentino:2020vhf}
E.~Di Valentino \textit{et al.},
``Snowmass2021 - Letter of interest cosmology intertwined I: Perspectives for the next decade,''
Astropart. Phys. \textbf{131}, 102606 (2021).


%\cite{Planck:2018vyg}
\bibitem{Planck:2018vyg}
N.~Aghanim \textit{et al.} [Planck Collaboration],
``Planck 2018 results. VI. Cosmological parameters,''
Astron. Astrophys. \textbf{641}, A6 (2020)
[erratum: Astron. Astrophys. \textbf{652}, C4 (2021)].


%\cite{Riess:2021jrx}
\bibitem{Riess:2021jrx}
A.~G.~Riess \textit{et al.},
``A Comprehensive Measurement of the Local Value of the Hubble Constant with 1 km/s/Mpc Uncertainty from the Hubble Space Telescope and the SH0ES Team,''
[arXiv:2112.04510 [astro-ph.CO]].

%\cite{Abdalla:2022yfr}
\bibitem{Abdalla:2022yfr}
E.~Abdalla \textit{et al.},
``Cosmology intertwined: A review of the particle physics, astrophysics, and cosmology associated with the cosmological tensions and anomalies,''
JHEAp \textbf{34}, 49-211 (2022).

%\cite{DiValentino:2020zio}
\bibitem{DiValentino:2020zio}
E.~Di Valentino \textit{et al.},
``Snowmass2021 - Letter of interest cosmology intertwined II: The hubble constant tension,''
Astropart. Phys. \textbf{131}, 102605 (2021).

%\cite{Verde:2019ivm}
\bibitem{Verde:2019ivm}
L.~Verde, T.~Treu and A.~G.~Riess,
``Tensions between the Early and the Late Universe,''
Nature Astron. \textbf{3}, 891 (2019).

%\cite{Knox:2019rjx}
\bibitem{Knox:2019rjx}
L.~Knox and M.~Millea,
``Hubble constant hunter\textquoteright{}s guide,''
Phys. Rev. D \textbf{101}, no.4, 043533 (2020).

%\cite{Jedamzik:2020zmd}
\bibitem{Jedamzik:2020zmd}
K.~Jedamzik, L.~Pogosian and G.~B.~Zhao,
``Why reducing the cosmic sound horizon alone can not fully resolve the Hubble tension,''
Commun. in Phys. \textbf{4}, 123 (2021).

%\cite{DiValentino:2021izs}
\bibitem{DiValentino:2021izs}
E.~Di Valentino \textit{et al.},
``In the realm of the Hubble tension\textemdash{}a review of solutions,''
Class. Quant. Grav. \textbf{38}, no.15, 153001 (2021).

%\cite{Perivolaropoulos:2021jda}
\bibitem{Perivolaropoulos:2021jda}
L.~Perivolaropoulos and F.~Skara,
``Challenges for \ensuremath{\Lambda}CDM: An update,''
New Astron. Rev. \textbf{95}, 101659 (2022).

%\cite{Shah:2021onj}
\bibitem{Shah:2021onj}
P.~Shah, P.~Lemos and O.~Lahav,
``A buyer\textquoteright{}s guide to the Hubble constant,''
Astron. Astrophys. Rev. \textbf{29}, no.1, 9 (2021).

%\cite{Kamionkowski:2022pkx}
\bibitem{Kamionkowski:2022pkx}
M.~Kamionkowski and A.~G.~Riess,
``The Hubble Tension and Early Dark Energy,''
[arXiv:2211.04492 [astro-ph.CO]].




%\cite{SDSS:2014iwm}
\bibitem{SDSS:2014iwm}
M.~Betoule \textit{et al.} [SDSS Collaboration],
``Improved cosmological constraints from a joint analysis of the SDSS-II and SNLS supernova samples,''
Astron. Astrophys. \textbf{568}, A22 (2014).


%\cite{Pan-STARRS1:2017jku}
\bibitem{Pan-STARRS1:2017jku}
D.~M.~Scolnic \textit{et al.} [Pan-STARRS1 Collaboration],
``The Complete Light-curve Sample of Spectroscopically Confirmed SNe Ia from Pan-STARRS1 and Cosmological Constraints from the Combined Pantheon Sample,''
Astrophys. J. \textbf{859}, no.2, 101 (2018).

%\cite{Scolnic:2021amr}
\bibitem{Scolnic:2021amr}
D.~M.~Scolnic \textit{et al.},
``The Pantheon+ Analysis: The Full Dataset and Light-Curve Release,''
[arXiv:2112.03863 [astro-ph.CO]].

%\cite{Brout:2022vxf}
\bibitem{Brout:2022vxf}
D.~Brout \textit{et al.},
``The Pantheon+ Analysis: Cosmological Constraints,''
[arXiv:2202.04077 [astro-ph.CO]].

%\cite{Beutler:2011hx}
\bibitem{Beutler:2011hx}
F.~Beutler \textit{et al.},
``The 6dF Galaxy Survey: Baryon Acoustic Oscillations and the Local Hubble Constant,''
Mon. Not. Roy. Astron. Soc. \textbf{416}, 3017-3032 (2011).

%\cite{Ross:2014qpa}
\bibitem{Ross:2014qpa}
A.~J.~Ross \textit{et al.},
``The clustering of the SDSS DR7 main Galaxy sample \textendash{} I. A 4 per cent distance measure at $z = 0.15$,''
Mon. Not. Roy. Astron. Soc. \textbf{449}, no.1, 835-847 (2015).

%\cite{BOSS:2016wmc}
\bibitem{BOSS:2016wmc}
S.~Alam \textit{et al.} [BOSS Collaboration],
``The clustering of galaxies in the completed SDSS-III Baryon Oscillation Spectroscopic Survey: cosmological analysis of the DR12 galaxy sample,''
Mon. Not. Roy. Astron. Soc. \textbf{470}, no.3, 2617-2652 (2017).

%\cite{Moresco:2020fbm}
\bibitem{Moresco:2020fbm}
M.~Moresco, R.~Jimenez, L.~Verde, A.~Cimatti and L.~Pozzetti,
``Setting the Stage for Cosmic Chronometers. II. Impact of Stellar Population Synthesis Models Systematics and Full Covariance Matrix,''
Astrophys. J. \textbf{898}, no.1, 82 (2020).
%\cite{Borghi:2021zsr}
%\bibitem{Borghi:2021zsr}
%N.~Borghi, M.~Moresco, A.~Cimatti, A.~Huchet, S.~Quai and L.~Pozzetti,
%``Toward a Better Understanding of Cosmic Chronometers: Stellar Population Properties of Passive Galaxies at Intermediate Redshift,''
%Astrophys. J. \textbf{927}, no.2, 164 (2022)
%\cite{Moresco:2016mzx}
%\bibitem{Moresco:2016mzx}
%M.~Moresco \textit{et al.},
%``A 6\% measurement of the Hubble parameter at $z\sim0.45$: direct evidence of the epoch of cosmic re-acceleration,''
%JCAP \textbf{05}, 014 (2016).

%\cite{DES:2017myr}
\bibitem{DES:2017myr}
T.~M.~C.~Abbott \textit{et al.} [DES Collaboration],
``Dark Energy Survey year 1 results: Cosmological constraints from galaxy clustering and weak lensing,''
Phys. Rev. D \textbf{98}, no.4, 043526 (2018).


%\cite{DES:2018ufa}
\bibitem{DES:2018ufa}
T.~M.~C.~Abbott \textit{et al.} [DES Collaboration],
``Dark Energy Survey Year 1 Results: Constraints on Extended Cosmological Models from Galaxy Clustering and Weak Lensing,''
Phys. Rev. D \textbf{99}, no.12, 123505 (2019).

%\cite{DES:2017qwj}
\bibitem{DES:2017qwj}
M.~A.~Troxel \textit{et al.} [DES Collaboration],
``Dark Energy Survey Year 1 results: Cosmological constraints from cosmic shear,''
Phys. Rev. D \textbf{98}, no.4, 043528 (2018).

%\cite{DES:2017hdw}
\bibitem{DES:2017hdw}
J.~Elvin-Poole \textit{et al.} [DES Collaboration],
``Dark Energy Survey year 1 results: Galaxy clustering for combined probes,''
Phys. Rev. D \textbf{98}, no.4, 042006 (2018).

%\cite{DES:2017gwu}
\bibitem{DES:2017gwu}
J.~Prat \textit{et al.} [DES Collaboration],
``Dark Energy Survey year 1 results: Galaxy-galaxy lensing,''
Phys. Rev. D \textbf{98}, no.4, 042005 (2018).

%\cite{Lewis:2013hha}
\bibitem{Lewis:2013hha}
A.~Lewis,
``Efficient sampling of fast and slow cosmological parameters,''
Phys. Rev. D \textbf{87}, no.10, 103529 (2013).

%\cite{Lewis:2019xzd}
\bibitem{Lewis:2019xzd}
A.~Lewis,
``GetDist: a Python package for analysing Monte Carlo samples,''
[arXiv:1910.13970 [astro-ph.IM]].




















\end{thebibliography}
\end{document}